\documentstyle[preprint,aps,floats]{revtex}

\begin{document}
\tightenlines
\renewcommand{\thefootnote}{\fnsymbol{footnote}}

\begin{flushright}
BI-TP 97/52, hep-ph/9711406\\
to appear in Nucl. Phys. B
\end{flushright}

\vspace{0.5cm}

\centerline{{\large \bf
Renormalization-scale-invariant continuation}}
\centerline{{\large \bf 
 of truncated QCD (QED) series -- }}
\centerline{{\large \bf 
an analysis beyond large-$\beta_0$ approximation}}

\vspace{1.cm}

\centerline{ G.~Cveti\v c\footnote[1]{e-mail:
cvetic@physik.uni-bielefeld.de; 
or: cvetic@doom.physik.uni-dortmund.de}}

\centerline{{\it  
Department of Physics, Universit\"at Bielefeld,
33501 Bielefeld, Germany}}

\renewcommand{\thefootnote}{\arabic{footnote}}

\begin{abstract}

An approximation algorithm is proposed to transform 
truncated QCD (or QED) series for observables. The approximation
is a modification of the Baker--Gammel approximants,
and is independent of the renormalization scale (RScl) $\mu$ 
-- the coupling parameter $\alpha(\mu)$ in the series and
in the resulting approximants can evolve according to 
the perturbative renormalization group equation (RGE) to any 
chosen loop order. The proposed algorithm is a natural 
generalization of the recently proposed method of diagonal 
Pad\'e approximants, the latter making the result RScl--invariant
in large-$\beta_0$ approximation for ${\alpha}(\mu)$.
The algorithm described below can extract large amount of 
information from a calculated available truncated perturbative 
series for an observable, by implicitly resumming large 
classes of diagrams.

\end{abstract}

\setcounter{equation}{0}
\newpage

\section{Introduction}
Pad\'e approximants (PA's) can be regarded as an improvement
to truncated perturbative series (TPS's), since they have the same
formal accuracy as the latter when expanded in powers of the
perturbative coupling parameter. In addition, PA's act as
a kind of analytical continuation to TPS's, and thus contain
explicitly some additional information not explicitly
(but implicitly) contained in the TPS's. PA's have been shown to
have interesting applications in statistical physics and
quantum field theory (QFT) \cite{SLS} and in QCD \cite{SEK}. In the latter
works \cite{SEK}, comparisons of the PA method with other methods which 
look for optimization of the TPS result via a judicial choice of the 
renormalization scale (RScl) and scheme (RSch) were performed
and showed good numerical agreement. These other methods include
the principle of minimal sensitivity ~\cite{Stevensonetal},
the BLM approach \cite{BLM} and its extensions
\cite{BLMext}--\cite{Neubertetal}, and
the effective charge approach~\cite{Grunberg}.
A novel RSch--invariant method, based partly
on the effective charge approach,
has recently been developed \cite{Maxwell}, and its precise relations
with the PA methods, as well as with the method presented here,
remain to be investigated. Another new approach \cite{Solovtsovetal}
reduces the RScl- and RSch-dependence by a method of
analytic continuation from the Euclidean to the time-like region
of a kinetic variable.
In this context, we mention
also a review of the role of power expansions in QFT
\cite{Fischer}. 

Recently, Gardi~\cite{Gardi} noted that the diagonal Pad\'e
approximants (dPA's) to truncated series for observables 
are invariant under the change of the renormalization scale
(RScl) $\mu$ in the large-$\beta_0$ approximation, i.e., when
the QCD coupling parameter ${\alpha}(\mu)$ is assumed to evolve
according to the one-loop renormalization group equation (RGE).
His observation was based on the invariance of dPA's
under the homographic transformations of the dPA 
argument (Ref.~\cite{BakerMorris}, Part I):
$z\!\mapsto\!az/(1+bz)$.
Since the (full knowledge of) observables must be
RScl--independent, Gardi's observation strongly suggested
that the dPA method correctly sums up certain classes of
multi--loop Feynman diagrams. Since the obtained dPA result
is invariant in the large--$\beta_0$ approximation,
it was conjectured that the summed-up diagrams are
one--gluon exchange diagrams where the gluon
propagator contains bubble--type of radiative corrections.
Later, it was explicitly shown~\cite{Brodskyetal} that
the latter conjecture in terms of Feynman diagrams
was correct. The authors of~\cite{Brodskyetal} stressed
that an extension of the approximation beyond the
large-$\beta_0$ approximation, i.e., beyond the dPA's,
remains a major outstanding question in the strive to
extract large amount of possible physical information from a
limited number of available (calculated) terms in the
formal QCD perturbation series of an observable.

In this paper, we present an algorithm which 
goes beyond the large--$\beta_0$ approximation.
It it based on a modified version of the diagonal
Baker--Gammel approximants (dBGA's)\footnote{
For the conventional Baker--Gammel approximants,
see for example Part II of Ref.~\cite{BakerMorris}.
See also later discussion.}
and results in an expression which has the following
properties:
\begin{enumerate}
\item it reproduces
the given truncated series up to any even order;
\item it is invariant under the change of the RScl $\mu$,
where the QCD coupling parameter $\alpha(\mu)$
appearing in the expression can be taken to evolve
under the RGE to any chosen loop order.
\end{enumerate}

The presented algorithm allows one to extract
a large amount of physical information -- in the sense of the
mentioned two points -- from a given QCD TPS for any
specific QCD observable if the latter
is available up to (and including) an even order.
However, it should be mentioned that the approximants of
the approach of the present paper, being relatively closely related 
to the usual PA's, may well not be able to discern in QCD
nonperturbative behavior arising from (UV and IR) renormalons -- see
discussions in Refs.~\cite{SEK}, \cite{Gardi}, \cite{Brodskyetal}
for the case of PA's. The latter behavior
is expected to manifest itselfs via a factorial growth of 
coefficients in TPS's.
The presented method may prove 
useful also in areas of physics other than QCD or QED.

\section{General method}
By scaling and raising to a power, it is possible
to redefine a generic QCD (or QED) observable $S$
in such a way that its (formal) perturbative series acquires 
the following dimensionless form
\begin{equation}
S \equiv z_1 f^{(1)}(z_1) = z_1 \left[
1 + r_1^{(1)} z_1 + r_2^{(1)} z_1^2 + \cdots 
+ r_n^{(1)} z_1^n + \cdots \right] \ ,
\label{S}
\end{equation}
where $z_1\!=\alpha(Q_1^2)/\pi$, $Q_1$ is a chosen renormalization
scale (RScl), and superscript in coefficients 
$r_n^{(1)}$ implies that they also depend
on the RScl $Q_1^2$: $r_n^{(1)}\!\equiv\!r_n(Q_1^2)$.
Since the RScl is chosen arbitrarily, the full result
$S$ is RScl--independent.\footnote{
However, $S$ does depend
on a certain energy scale $Q$ typical for the process involved.
In scattering processes,
$Q$ is of the order of the center--of--mass energy.}
However, as a rule, only a very limited number of coefficients
$r_j^{(1)}$ are available (calculated), i.e., only a truncated
perturbation series
\begin{equation}
S_n^{(1)} \equiv z_1 f^{(1)}(z_1;n) = z_1 \left[
1 + r_1^{(1)} z_1 + r_2^{(1)} z_1^2 + \cdots 
+ r_n^{(1)} z_1^n \right]
\label{Sn}
\end{equation}
is available, and it is therefore explicitly RScl--dependent
as indicated by the superscripts: $S_n^{(1)}\!\equiv\!S_n(Q_1^2)$.
Hence, there appears naturally the question of how to
construct, on the basis of (\ref{Sn}), an expression
which satisfies the two previously mentioned points.
We will now present an algorithm for constructing just
such approximations to $S^{(1)}_n$'s (and therefore to $S$),
and subsequently show with two theorems and their proofs that
the constructed approximations satisfy the two mentioned points.

The evolving QCD (or QED) coupling parameter $\alpha$ satisfies
the following RGE:
\begin{equation}
\frac{dx}{dt} = - \sum_{j=0}^{\infty} {\beta}_j x^{j+2} 
\qquad \left[ x \equiv {\alpha}(p^2)/\pi, \ dt \equiv d \ln p^2 \right] \ , 
\label{alphaRGE}
\end{equation}
where $\beta_j$ are constants (RScl--independent),
and $\beta_0$ and $\beta_1$ are even 
renormalization--scheme--independent (RSch--independent).
We can choose to take into account any number of terms on the
right of this RGE. Now we define the quantity
\begin{equation}
k(z_1,u_1) \equiv \frac{ {\alpha}(p^2) }{ {\alpha}(Q_1^2) }
\qquad
\left[ z_1 = {\alpha}(Q_1^2)/\pi, \ u_1=\ln(p^2/Q_1^2)\right] . \ 
\label{kdef}
\end{equation}
Its formal Taylor expansion in powers
of $u_1\!\equiv\!\ln(p^2/Q_1^2)$ is
\begin{equation}
k(z_1,u_1) = 1 + \sum_{j=1}^{\infty} u_1^j k_j(z_1) \ ,
\qquad \mbox{where: } \ k_j(z_1) = \frac{1}{j!} 
\frac{\partial^j}{\partial u^j} k(z_1, u) {\big |}_{u=0} \ .
\label{kTaylor}
\end{equation}
On the basis of RGE (\ref{alphaRGE}), we have
\begin{equation}
k_j(z_1) = (-1)^j \beta_0^j z_1^j + {\cal O}(z_1^{j+1}) \ , \quad
k_0(z_1) = 1 \ .
\label{kjsapprox}
\end{equation}
Now we simply rearrange the formal series (\ref{S}) for $S/z_1$,
which is in powers of $z_1\!\equiv\!\alpha(Q_1^2)/\pi$,
into a related series in $k_j(z_1)$
\begin{equation}
S = z_1 \left[ 1 + \sum_{j=1}^{\infty} f_j^{(1)} k_j(z_1) \right] \ .
\label{Sinkj}
\end{equation} 
The superscript in the coefficients $f_j^{(1)}$ now again means
that they are functions of the RScl $Q_1^2$: 
$f_j^{(1)}\!\equiv\!f_j(Q_1^2)$.
We now define the corresponding formal series ${\cal F}^{(1)}$ 
in powers of $(-z_1)$
\begin{equation}
z_1 {\cal F}^{(1)}(z_1) \equiv z_1 \left[
1 + \sum_{j=1}^{\infty} f_j^{(1)} (-z_1)^j \right] \ ,
\label{F1}
\end{equation} 
and construct for $z_1 {\cal F}^{(1)}(z_1)$ the
diagonal Pad\'e approximants (dPA's)
\begin{eqnarray}
z_1 [ M-1 / M ]_{ {\cal F}^{(1)} }(z_1) &=&
z_1 \left[ 1 + \sum_{m=1}^{M-1}a_m z_1^m \right]
\left[ 1 + \sum_{n=1}^{M}b_n z_1^n \right]^{-1} \ ,
\label{PAF11}
\\
z_1 {\cal F}^{(1)}(z_1) &=&
z_1 [ M-1 / M ]_{ {\cal F}^{(1)} }(z_1) + 
{\cal O} \left( z_1^{2M+1} \right) \ .
\label{PAF12}
\end{eqnarray}
We recall that $f_j^{(1)}$ is
a function of only $r_1^{(1)},\!\ldots,\!r_j^{(1)}$, due to
relations (\ref{kjsapprox}).
Therefore, also the above dPA, depending only on
$f_i^{(1)}$ ($i\!=\!1,\!\ldots,\!2M\!-\!1$) by (\ref{PAF12}), 
depends only on the first $2M\!-\!1$ coefficients $r_i^{(1)}$
($i\!=\!1,\!\ldots,\!2M\!-\!1$) of the original series (\ref{S}).
At this point of algorithm, we perform decomposition of the
dPA (\ref{PAF11}) into simple fractions
\begin{equation}
z_1 [ M-1 / M ]_{ {\cal F}^{(1)} }(z_1) =
z_1 \sum_{i=1}^M \frac{ {\tilde \alpha}_i }
{ ( 1 + {\tilde u}_i z_1 ) } \ ,
\label{PAF1decomp}
\end{equation}
where $[- 1/{\tilde u}_i]$ are the $M$ zeros of the
denominator polynomial of the dPA (\ref{PAF11}). As long as the latter
polynomial doesn't have multiple zeros,\footnote{
We assume that the exceptional situation of multiple zeros 
of the denominator does not appear.
See also later discussion following Eq.~(\ref{Sint}).} 
this decomposition is possible and unique. Although the
sum (\ref{PAF1decomp}) is real for real $z_1$,
parameters ${\tilde u}_i$ and ${\tilde \alpha}_i$ are 
in general complex numbers. They
depend on the RScl $Q_1^2$, because they are
functions of $r_1^{(1)},\!\ldots,\!r_{2M-1}^{(1)}$ which
themselves are functions of $Q_1^2$. 
Now we come to the central part of the algorithm, by
constructing a modified version of the diagonal Baker--Gammel
approximants (dBGA's) to the observable 
$S = z_1 f^{(1)}$ of Eq.~(\ref{S})
\begin{equation}
z_1 G_{ f^{(1)} }^{ [M-1/M] }(z_1) \equiv 
z_1 \sum_{i=1}^{M} {\tilde \alpha}_i k ( z_1,{\tilde u}_i ) \ .
\label{dBGAdef}
\end{equation}
Function $k(z,u)$, which depends on two arguments and is
in our case defined via (\ref{kdef}) and (\ref{alphaRGE}),
can be called the kernel function of the above modified
dBGA.
As mentioned previously, the dBGA (\ref{dBGAdef}) is uniquely
determined by the first $(2M\!-\!1)$ coefficients $r^{(1)}_i$
($i\!=\!1,\!\ldots,\!2M\!-\!1$) of the original perturbative
series (\ref{S}), i.e., knowledge of the truncated series
$S_{2M-1}$ (\ref{Sn}) uniquely determines (\ref{dBGAdef}).
Now we show, by proving two theorems, that these modified dBGA's
have precisely the two wanted properties mentioned earlier on.

{\em Theorem 1 (``Approximation'' theorem):\/} The modified dBGA
(\ref{dBGAdef}) of order $(2M\!-\!1)$
approximates the observable $S\!\equiv\!z_1f^{(1)}(z_1)$ 
of Eq.~(\ref{S}) up to (and including) ${\cal O}(z_1^{2M})$:
\begin{equation}
S = z_1 G_{ f^{(1)} }^{ [M-1/M] }(z_1) + 
{\cal O} \left( z_1^{2M+1} \right) \ .
\label{Theor1}
\end{equation} 

{\em Proof:\/}  We use for $k( z_1, {\tilde u}_i )$ in
the definition (\ref{dBGAdef}) of the modified dBGA the formal Taylor 
expansion (\ref{kTaylor}) in powers of ${\tilde u}_i$
\begin{equation}
z_1 G_{ f^{(1)} }^{ [M-1/M] }(z_1) =
z_1 \sum_{i=1}^M {\tilde \alpha}_i 
\sum_{m=0}^{\infty} {\tilde u}_i^m k_m(z_1) =
z_1 \sum_{m=0}^{\infty} k_m(z_1) \left[ \sum_{i=1}^M
{\tilde \alpha}_i {\tilde u}_i^m \right] \ .
\label{expdBGA}
\end{equation}
Here we implicitly assume convergence of the above series.
We do the same kind of formal expansion for the corresponding dPA
(\ref{PAF11})--(\ref{PAF1decomp})
\begin{equation}
z_1 [M-1/M]_{ {\cal F}^{(1)} } (z_1) =
z_1 \sum_{i=1}^M {\tilde \alpha}_i 
\sum_{m=0}^{\infty} {\tilde u}_i^m (-z_1)^m =
z_1 \sum_{m=0}^{\infty} (-z_1)^m \left[ \sum_{i=1}^M
{\tilde \alpha}_i {\tilde u}_i^m \right] \ .
\label{expdPA}
\end{equation}
According to (\ref{PAF12}), this expression reproduces
(``approximates'') the series (\ref{F1}) up to, and including,
the term $z_1 f^{(1)}_{2M-1} (-z_1)^{2M-1}\!\sim\!z_1^{2M}$.
Therefore,
\begin{equation}
\sum_{i=1}^M {\tilde \alpha}_i {\tilde u}_i^m = f^{(1)}_m
\qquad (m=0,1,\ldots,2M-1) \ .
\label{coeffs}
\end{equation}
Using this in (\ref{expdBGA}) and comparing with the full
formal series (\ref{Sinkj}) for the observable
$S\!\equiv\!z_1 f^{(1)}(z_1)$, we obtain
\begin{eqnarray}
z_1 G_{ f^{(1)} }^{ [M-1/M] }(z_1) &=&
z_1 \sum_{m=0}^{2M-1} f^{(1)}_m k_m(z_1) + 
{\cal O} \left( z_1 k_{2M}(z_1) \right) \ \Longrightarrow
\label{dBGAapprox1}
\\
S - z_1 G_{ f^{(1)} }^{ [M-1/M] }(z_1) &=&
z_1 \sum_{m=2M}^{\infty} f^{(1)}_m k_m(z_1) = 
{\cal O} \left( z_1 k_{2M}(z_1) \right) =
{\cal O} \left( z_1^{2M+1} \right) \ ,
\label{dBGAapprox2}
\end{eqnarray}
where at the end we used estimates (\ref{kjsapprox})
[$k_j(z_1)\!\sim\!z_1^j$]. Relation (\ref{dBGAapprox2})
proves the theorem. We stress, however, that the formal series
involved in the proof
were implicitly assumed to be convergent. The question
of when the proof survives once we abandon
the assumption of convergence is left open.

{\em Theorem 2 (Invariance under argument transformation):\/}
The modified diagonal Baker--Gammel approximant
(dBGA) $z_1 G_{ f^{(1)} }^{ [M-1/M] }(z_1)$,
as defined in Eq.~(\ref{dBGAdef}),
for the observable $S\!\equiv\!z_1 f^{(1)}(z_1)$
of Eq.~(\ref{S}), is invariant under the following transformations
of the argument:
\begin{equation}
z_1 \mapsto z_2(z_1; u_{21}) \quad
\mbox{such that:} \ z_1 k(z_1,u_1)=z_2 k(z_2,u_1-u_{21}) \ ,
\label{Theor21}
\end{equation}
where $u_{21}$ is any (arbitrary) fixed complex number,
$u_1$ is arbitrary complex number, $z_2( z_1; u_{21})$
is independent of $u_1$, and the following relations
are assumed for the $k$-function (and its Taylor coefficients)
appearing in the dBGA:
\begin{equation}
k(z,u) \sim z^0 (=1) \ , \quad 
k_j(z) \sim z^j \ (\Rightarrow z_1 \sim z_2) \ .
\label{Theor22}
\end{equation}

{\em Note to Theorem 2:\/} The seemingly artificial and
complicated set of conditions for argument transformation
is motivated by the following QCD (or QED) interpretation of the
above parameters:
\begin{equation}
z_1 \equiv \frac{\alpha(Q_1^2)}{\pi} \ , \quad
u_1 = \ln \left( \frac{p^2}{Q_1^2} \right) \ , \quad
k(z_1,u_1) = \frac{ {\alpha}(p^2) }{ {\alpha}(Q_1^2) } \ , \quad
u_{21} = \ln \left( \frac{Q_2^2}{Q_1^2} \right) \ .
\label{comm1Theor2}
\end{equation}
We see then from here and from (\ref{kjsapprox}) that
conditions (\ref{Theor22}) for Theorem 2 are fulfilled in the
QCD (or QED) case and that the argument transformation in this case
means simply the change of the renormalization scale (RScl)
from $Q_1$ to $Q_2$
\begin{equation}
z_1 \equiv \frac{ {\alpha}(Q_1^2) }{\pi}  \mapsto  
z_2 \equiv \frac{ {\alpha}(Q_2^2) }{\pi} \ , \quad
(u_1-u_{21}) = \ln \left( \frac{ p^2 }{ Q_2^2 } \right) 
\left( \equiv u_2 \right) \ , \quad
k(z_2,u_2)= \frac{ {\alpha}(p^2) }{ {\alpha}(Q_2^2) } \ ,
\label{comm2Theor2}
\end{equation}
thus implying RScl--invariance of the dBGA (\ref{dBGAdef}).

{\em Proof of Theorem 2:\/}
Let $z_1$ and $z_2$ be related by transformation (\ref{Theor21}).
We have two different formal series of $S$ in $k_j(z_1)$ and $k_j(z_2)$
\begin{equation}
S \left( = z_1 f^{(1)}(z_1) \right)
= z_1 \left[ 1 + \sum_{j=1}^{\infty} f^{(1)}_j k_j(z_1) \right] =
z_2 \left[ 1 + \sum_{j=1}^{\infty} f^{(2)}_j k_j(z_2) \right] 
\left( = z_2 f^{(2)}(z_1) \right) \ .
\label{Sz1z2}
\end{equation}
[In QCD or QED: $f^{(1)}_j\!\equiv\!f_j(Q_1^2)$ and
$f^{(2)}_j\!\equiv\!f_j(Q_2^2)$.]
The corresponding two formal series in simple powers of $z_1$ and $z_2$,
respectively, are defined in the described algorithm by (\ref{F1})
\begin{equation}
z_1 {\cal F}^{(1)}(z_1) \equiv z_1 \left[
1 + \sum_{j=1}^{\infty} f_j^{(1)} (-z_1)^j \right] \ , \quad
z_2 {\cal F}^{(2)}(z_2) \equiv z_2 \left[
1 + \sum_{j=1}^{\infty} f_j^{(2)} (-z_2)^j \right] \ .
\label{Fz1z2}
\end{equation}
Following the algorithm, we construct the usual dPA's for 
$z_1 {\cal F}^{(1)}(z_1)$ and $z_2 {\cal F}^{(2)}(z_2)$ in the
form of decomposition into simple fractions (\ref{PAF1decomp})
\begin{equation}
z_1 [ M-1 / M ]_{ {\cal F}^{(1)} }(z_1) =
z_1 \sum_{i=1}^M \frac{ {\tilde \alpha}_i }
{ ( 1 + {\tilde u}_i z_1 ) } \ , \quad
z_2 [ M-1 / M ]_{ {\cal F}^{(2)} }(z_2) =
z_2 \sum_{i=1}^M \frac{ {\tilde {\tilde \alpha}}_i }
{ ( 1 + {\tilde {\tilde u}}_i z_2 ) } \ ,
\label{dPAF12dec}
\end{equation}
again assuming that the exceptional case of multiple poles
doesn't appear. The modified dBGA's are then constructed
in both cases according to the algorithm, by Eq.~(\ref{dBGAdef})
\begin{equation}
z_1 G_{ f^{(1)} }^{ [M-1/M] }(z_1) \equiv 
z_1 \sum_{i=1}^{M} {\tilde \alpha}_i k ( z_1,{\tilde u}_i ) \ ,
\quad
z_2 G_{ f^{(2)} }^{ [M-1/M] }(z_2) \equiv
z_2 \sum_{i=1}^{M} {\tilde {\tilde \alpha}}_i 
k ( z_2,{\tilde {\tilde u}}_i ) \ .
\label{dBGA12}
\end{equation}
We will now show that 
${\tilde {\tilde \alpha}}_i\!=\!{\tilde \alpha}_i$ and
${\tilde {\tilde u}}_i\!=\!{\tilde u}_i-u_{21}$.
Define expression
\begin{equation}
z_2 {\tilde {\cal G}}_{ f^{(2)} }^{ [M-1/M] }(z_2) \equiv 
z_2 \sum_{i=1}^{M} {\tilde \alpha}_i k ( z_2,{\tilde u}_i-u_{21} ) \ .
\label{dBGArz2}
\end{equation}
Transformation (\ref{Theor21}) $z_1\!\mapsto\!z_2$
and definitions (\ref{dBGA12}) and (\ref{dBGArz2}) imply
\begin{equation}
z_2 {\tilde {\cal G}}_{ f^{(2)} }^{ [M-1/M] }(z_2) =
z_1 G_{ f^{(1)} }^{ [M-1/M] }(z_1) \ .
\label{dBGAz1z2}
\end{equation}
Therefore, by Theorem 1 [Eq.~(\ref{Theor1})] we have
\begin{equation}
S - z_2 {\tilde {\cal G}}_{ f^{(2)} }^{ [M-1/M] }(z_2) =
S - z_1 G_{ f^{(1)} }^{ [M-1/M] }(z_1) \sim
z_1^{2M+1} \sim  z_2^{2M+1} \sim
 z_2 k_{2M}(z_2) \ ,
\label{dBGAapprox}
\end{equation}
where at the end we used conditions (\ref{Theor22}). We now make 
the formal Taylor expansion of (\ref{dBGArz2}) in powers of
$({\tilde u}_i-u_{21})$
\begin{equation}
z_2 {\cal G}_{ f^{(2)} }^{ [M-1/M] }(z_2) =
z_2 \sum_{i=1}^M {\tilde \alpha}_i 
\sum_{m=0}^{\infty} \left( {\tilde u}_i - u_{21} \right)^m k_m(z_2) =
z_2 \sum_{m=0}^{\infty} k_m(z_2) \left[ \sum_{i=1}^M
{\tilde \alpha}_i \left( {\tilde u}_i - u_{21} \right)^m \right] \ .
\label{expdBGArz2}
\end{equation}
Comparing (\ref{dBGAapprox}) and (\ref{expdBGArz2}), 
and using (\ref{Sz1z2}) for the case of argument $z_2$, 
we obtain relations analogous to those
for the case of argument $z_1$ (\ref{coeffs})
\begin{equation}
\sum_{i=1}^M {\tilde \alpha}_i \left( {\tilde u}_i - u_{21} \right)^m 
= f^{(2)}_m
\qquad (m=0,1,\ldots,2M-1) \ .
\label{coeffsz2}
\end{equation}
Therefore, the dPA for $z_2 {\cal F}^{(2)}(z_2)$ of (\ref{Fz1z2}) is
\begin{equation}
z_2 [ M-1 / M ]_{ {\cal F}^{(2)} }(z_2) =
z_2 \sum_{i=1}^M \frac{ {\tilde \alpha}_i }
{ \left[ 1 + \left({\tilde u}_i-u_{21} \right) z_2 \right]  } =
z_2 \sum_{m=0}^{\infty} (-z_2)^m \left[ \sum_{i=1}^M
{\tilde \alpha}_i \left( {\tilde u}_i - u_{21} \right)^m \right] \ ,
\label{dPArz2}
\end{equation}
since this dPA expression reproduces the power terms of
$z_2 {\cal F}^{(2)}(z_2)$ up to (and including) $\sim\!z_2^{2M}$,
as can be seen from explicit expansions (\ref{Fz1z2}), (\ref{dPArz2})
and relations (\ref{coeffsz2})
\begin{equation}
z_2 {\cal F}^{(2)}(z_2) -
z_2 \sum_{i=1}^M \frac{ {\tilde \alpha}_i }
{ \left[ 1 + \left({\tilde u}_i-u_{21} \right) z_2 \right]  } 
= {\cal O} \left( z_2^{2M+1} \right) \ .
\label{dPAz2approx}
\end{equation}
Therefore, this shows that (\ref{dBGArz2}) is in fact the modified
dBGA $z_2 G_{ f^{(2)} }^{ [M-1/M] } (z_2)$ (\ref{dBGA12})
to the observable $S$ for the case of argument $z_2$,
according to the described algorithm of constructing the modified dBGA's
(\ref{dBGAdef})
\begin{equation}
\left[ z_2 {\tilde {\cal G}}_{ f^{(2)} }^{ [M-1/M] }(z_2) = \right]
z_2 \sum_{i=1}^M {\tilde \alpha}_i k(z_2, {\tilde u}_i-u_{21}) =
z_2 G_{ f^{(2)} }^{ [M-1/M] } (z_2) \ .
\label{dBGAz2eq}
\end{equation}
This, together with equality (\ref{dBGAz1z2}), shows that the
modified dBGA of order $(2M\!-\!1)$, as defined by the described 
algorithm leading to
(\ref{dBGAdef}), is really invariant under transformations
of argument (\ref{Theor21}).
Thus the proof of the theorem is completed.

Again, it should be emphasized that it is left open
under which circumstances the proof remains valid once we abandon
the assumption of convergence of the series involved.

In the described approach,
the special QCD (or QED) case of 
one--loop evolution of ${\alpha}(p^2)$
(large-${\beta}_0$ approximation)
means: $k(z_1,u_1)\!=\!1/(1\!+\!{\beta}_0z_1u_1)$, 
$k_j(z)\!=\!{\beta}_0^j (-z_1)^j$. Therefore, in the one--loop
case, expansion (\ref{Sinkj}) for $S/z_1$ and (\ref{F1})
for ${\cal F}^{(1)}(z_1)$ are identical if $z_1$ in the latter
series is replaced by ${\beta}_0 z_1$ (rescaling). The
dBGA (\ref{dBGAdef}) is in this case reduced to the usual dPA
(\ref{PAF1decomp}), with $z_1$ in the denominators replaced by
${\beta}_0 z_1$.

The quantity $S/z_1\!\equiv\!f^{(1)}(z_1)$ of (\ref{S})
can lead, at least in some cases, to nonmodified
dBGA's, i.e., those for which the series (\ref{F1})
has the special property of being a
Hamburger series (cf.~Ref.~\cite{BakerMorris}, Part II).\footnote{
We note that authors of \cite{BakerMorris} discuss a
somewhat more constrained version of nonmodified (d)BGA's,
i.e., the one for which ${\cal F}^{(1)}(z_1)$ is
a Stieltjes series, corresponding to expression
(\ref{Hambfun}) with lower integration bound being zero.} 
This series is the formal Taylor expansion of a
Hamburger function
\begin{equation}
{\cal F}^{(1)}(z_1) = \int_{-\infty}^{\infty}
\frac{ d {\phi}(u_1) }{1 + u_1 z_1} \ ,
\label{Hambfun}
\end{equation}
where ${\phi}(u_1)$ is increasing with increasing $u_1$,
and the coefficients
$f_j^{(1)}$ in (\ref{F1}) are the finite moments
\begin{equation}
f_j^{(1)} = \int_{-\infty}^{\infty} u_1^j d {\phi}(u_1) \qquad
(j=0,1,2,\ldots) \ , \qquad (f_0^{(1)}=1) \ .
\label{momHamb}
\end{equation}
In this special case, observable $S$ of (\ref{S})
has the integral representation
\begin{equation}
S = z_1 f^{(1)}(z_1) = \int_{-\infty}^{\infty}
z_1 k(z_1,u_1) d {\phi}(u_1) \ .
\label{Sint}
\end{equation}
In such (special) cases, the parameters in decomposition 
(\ref{PAF1decomp}) of the dPA's are real (${\tilde \alpha}_i$'s
are positive)~\cite{BakerMorris}, and the resulting
dBGA's (\ref{dBGAdef}) are therefore manifestly real
numbers. Decomposition (\ref{PAF1decomp}) is in this case
always possible, i.e., the dPA (\ref{PAF11}) has no multiple zeros.
The case of (\ref{Sint}), reinterpreted
in terms of QCD parameters (\ref{comm1Theor2}), 
means that in the integral for $S$ (over $du_1\!\equiv\!d p^2/p^2$)
there is a positive function
$\rho(u_1)\!=\!{\phi}^{\prime}(u_1)$. The latter
can be interpreted as momentum probability
distribution \cite{Neubertetal} (see also   
\cite{LMCF} and \cite{Brodskyetal}) in diagrams with
exchange of one effective virtual gluon,
where the gluonic propagator contains radiative corrections.
In fact, if the (approximate) $S$ contains only contributions
which can be reduced to such classes of one--gluon--exchange 
diagrams, the authors of \cite{Brodskyetal} conjectured that 
function $\rho(u_1)$ is then positive definite.
Their conjecture would therefore suggest
the following: in the cases where the truncated series
$S^{(1)}_n$ of (\ref{Sn}) contains only contributions of
diagrams with at most one gluon exchange in
each one--particle--irreducible part,
the method of the modified dBGA's presented
here would result in a Hamburger series
for ${\cal F}^{(1)}$ of (\ref{F1}) and hence in
nonmodified dBGA's, i.e., with ${\tilde \alpha}_i$ and ${\tilde u}_i$
in (\ref{dBGAdef}) being real (${\tilde \alpha}_i\!>\!0$).
Having a truncated series $S^{(1)}_n$ of (\ref{Sn}), 
with $n\!=\!2M-1$ odd, it is
straightforward to check whether such a case sets in
(cf.~Part I of \cite{BakerMorris}): determinants
of a limited set of matrices $A[0],\!\ldots,\!A[(n\!-\!1)/2]$ 
have to be positive, where $A[m]$ is a 
$(m\!+\!1)\!\times\!(m\!+\!1)$ matrix whose elements are
$A_{ij}[m]\!\equiv\!f^{(1)}_{i+j-2}$
($i,\!j\!=\!1,\!\ldots,\!m\!+1$).
For example, for $n\!=\!1$ ($S^{(1)}_1$) 
this case always sets in, 
and for $n\!=\!3$ ($S^{(1)}_3$) it sets in when\footnote{
See also explicit discussion of $n\!=\!1$ ($M\!=\!1$) and
$n\!=\!3$ ($M\!=\!2$) cases later on.}
$f_2^{(1)}\!-\!(f_1^{(1)})^2\!>\!0$.

We decided to use for the result (\ref{dBGAdef}) the
terminology ``modified diagonal Baker-Gammel approximant'' 
primarily due to the previously mentioned
indirect connection of our result with the nonmodified 
BGA's discussed in \cite{BakerMorris}. The latter
reference (part II, Chapter 1.2) briefly discusses
BGA's, for some specific kernels $k(z,u)$, and from the
mathematical and numerical point of view. Theorem 1
of the present paper is indirectly related to the
Convergence Theorem 1.2.1 of part II of Ref.~\cite{BakerMorris}
for BGA's, while Theorem 2 of the present paper
(RScl--invariance) has no analog in \cite{BakerMorris}.
We stress that the presentation of the present paper
is self--contained in the sense that the reader is not
required to know anything about BGA's, but should 
be reasonably familiar with the usual (d)PA's.
In the previous paragraph (which does not represent
an essential part of the paper), 
some familiarity with the Hamburger series 
was assumed.

What to do when parameters
${\tilde \alpha}_i$ and ${\tilde u}_i$ in the
modified dBGA (\ref{dBGAdef}) are not simultaneously real?
In that case, modified dBGA could in principle be complex.
However, Theorems 1 and 2 are valid not just for the
entire modified dBGA's, but also for their real
(and imaginary) parts. Since the observable $S$
is real, we then just take the real part of expression
(\ref{dBGAdef}). In fact, since $z_1$ and $S$ are real,
Theorem 1 [Eq.~(\ref{Theor1})] shows that the imaginary part
of the modified dBGA $z_1 G_{ f^{(1)} }^{ [M-1/M] }(z_1)$
must be $\sim\!z_1^{2M+1}$ or even less.

\section{Explicit examples}
It is instructive to obtain explicit formulas for constructing
the dBGA's $z_1 G_{ f^{(1)} }^{[M-1/M]}(z_1)$ of (\ref{dBGAdef})
once we know the truncated perturbative QCD (or QED) series
$S^{(1)}_{2M-1}$ of Eq.~(\ref{Sn})
for the practically possible cases of $M\!=\!1$ or $M\!=\!2$.
This will be done below.

The case $M\!=\!1$ in QCD (or QED) is the case investigated already by
Brodsky, Lepage and Mackenzie (BLM)\cite{BLM}. Formalism presented
here, under the mentioned QCD (or QED) identifications
(\ref{comm1Theor2}) and (\ref{alphaRGE}) for the parameters,
gives
\begin{equation}
S^{(1)}_1 \equiv z_1 f^{(1)}(z_1;1) = z_1 \left[ 1+r^{(1)}_1 z_1 \right]
\ \Rightarrow \ 
z_1 G_{ f^{(1)} }^{[0/1]}(z_1) = 
z_1 k(z_1, -r^{(1)}_1/{\beta}_0) \ .
\label{BLMrel}
\end{equation}
We note that the above approximation, according to
notation (\ref{comm1Theor2}), is in fact the value of
the coupling parameter at a scale $Q^2$
\begin{equation}
z_1 G_{ f^{(1)} }^{[0/1]}(z_1)  = {\alpha}(Q^2)/{\pi} \ ,
\quad Q^2 = Q_1^2 \exp( -r^{(1)}_1/{\beta}_0 ) \ .
\label{BLM2rel}
\end{equation}
We note that ${\alpha}(p^2)$ evolves according to RGE
(\ref{alphaRGE}) where the number of retained terms (loops)
can be arbitrarily chosen if known.
The result $z_1 G_{ f^{(1)} }^{ [0/1] }(z_1)$
and the scale $Q^2$ are independent of the choice 
of the RScl $Q_1^2$ (up to the chosen loop-order)
according to Theorem 2, 
while $S_1^{(1)}$[$\equiv\!S_1(Q_1^2)$] and
$S_1^{(2)}$[$\equiv\!S_1(Q_2^2)$] in general differ for
RScl's $Q_1^2\!\not=\!Q_2^2$: $S_1^{(2)}\!-\!S_1^{(1)}\!\sim\!z_1^3$.
In fact, it is straightforward to check directly that
$z_1 G_{ f^{(1)} }^{ [0/1] }(z_1)\!=\!z_2 G_{ f^{(2)} }^{ [0/1] }(z_2)$,
i.e., that $Q^2$ of (\ref{BLM2rel}) is RScl--independent, because
$r^{(2)}_1\!-\!r^{(1)}_1\!\equiv\!r_1(Q_2^2)\!-\!r_1(Q_1^2)\!=\!{\beta}_0 
\ln (Q_2^2/Q_1^2)\!\not=\!0$ as can be checked by using (\ref{alphaRGE}).
This result can be formulated also in a more intuitive manner: 
when choosing the RScl equal to $Q^2$ of (\ref{BLM2rel})
[$z_1\!\equiv\!z(Q_1^2)\!\mapsto\!z(Q^2)$, 
$r^{(1)}_1\!\equiv\!r_1(Q_1^2)\!\mapsto\!r_1(Q^2)$,
$S^{(1)}_1\!\equiv\!S_1(Q_1^2)\!\mapsto\!S_1(Q^2)$],
the next-to-leading (NLO) term [$\sim\!z^2(Q^2)$]
in the perturbative expansion of $S$ is zero:
$S\!=\!z(Q^2) [ 1\!+\!{\cal O}(z^2(Q^2))]$. This follows
from (\ref{BLM2rel}) (i.e., implicitly from RScl--invariance
Theorem 2 for $M\!=\!1$ case)  and from Theorem 1 [Eq.~(\ref{Theor1})],
under inclusion of conditions (\ref{Theor22}) [$z_1\!\sim\!z(Q^2)$].

The case $M\!=\!2$ in QCD (or QED) is 
algebraically more involved, but can be
worked out in a straightforward way using the presented formalism.
Expansion functions $k_j(z_1)$ (\ref{kTaylor}) are obtained from
RGE (\ref{alphaRGE})
\begin{eqnarray}
k_1(z_1) & = & - {\beta}_0 z_1 - {\beta}_1 z_1^2 - {\beta}_2 z_1^3 
- \ldots \ ,
\label{k1}
\\
k_2(z_1) &=& + {\beta}_0^2 z_1^2 + (5/2) {\beta}_0 {\beta}_1 z_1^3 +
\ldots \ , \qquad k_3(z_1) = - {\beta}_0^3 z_1^3 - \ldots \ .
\label{k2k3}
\end{eqnarray}
Inverting these relations gives
\begin{eqnarray}
z_1^3 &=& - \frac{1}{ {\beta}_0^3 } k_3(z_1) + {\cal O}(k_4) \ ,
\qquad z_1^2 = + \frac{1}{ {\beta}_0^2 } k_2(z_1) +
\frac{5 {\beta}_1 }{2 {\beta}_0^4 } k_3(z_1) + {\cal O}(k_4) \ ,
\label{z1312}
\\
z_1 &=&  - \frac{1}{ {\beta}_0 } k_1(z_1) -
\frac{ {\beta}_1 }{ {\beta}_0^3 } k_2(z_1) -
\left( \frac{ 5 {\beta}_1^2 }{ 2 {\beta}_0^5 }
- \frac{ {\beta}_2 }{ {\beta}_0^4 } \right)  k_3(z_1)
+ {\cal O}(k_4) \ .
\label{z11}
\end{eqnarray}
Inserting this into the truncated series
$S^{(1)}_3$ of Eq.~(\ref{Sn}) (presumed available), 
results in the rearranged truncated series for $S^{(1)}_3$ of the
form (\ref{Sinkj}), with coefficients
\begin{eqnarray}
f^{(1)}_1 &=& - \frac{r^{(1)}_1 }{ {\beta}_0 } \ ,
\qquad
f^{(1)}_2 = - \frac{ {\beta}_1 }{ {\beta}_0^3 } r^{(1)}_1
+ \frac{1}{ {\beta}_0^2 } r^{(1)}_2 \ ,
\label{f11f12}
\\
f^{(1)}_3 &=& \left( - \frac{ 5 {\beta}_1^2 }{ 2 {\beta}_0^5 }
+ \frac{ {\beta}_2 }{ {\beta}_0^4 } \right)  r^{(1)}_1
+ \frac{ 5 {\beta}_1 }{ 2 {\beta}_0^4 } r^{(1)}_2
- \frac{1}{ {\beta}_0^3 } r^{(1)}_3 \ .
\label{f13}
\end{eqnarray}
Having these coefficients, we construct the truncated series
for $z_1 {\cal F}^{(1)}$ (\ref{F1}), and the dPA
$z_1 [1/2]_{ {\cal F}^{(1)} }(z_1)$ (\ref{PAF12}) in its decomposed form
(\ref{PAF1decomp}). The resulting expressions for
parameters ${\tilde u}_i$ and ${\tilde \alpha}_i$ are
\begin{eqnarray}
{\tilde u}_{2,1} &=& \left[ (f_3-f_1 f_2) \pm \sqrt{ {\rm det} } \right]
\left[ 2 (f_2 - f_1^2) \right]^{-1} \ ,
\label{tildeu21}
\\
\mbox{where: } \
{\rm det} &=& \left[ f_3 + f_1 (2 f_1^2 - 3 f_2) \right]^2 +
4 (f_2 - f_1^2)^3 \ ,
\label{det}
\\
{\tilde \alpha}_1 &=& ( {\tilde u}_2 - f_1 )/
( {\tilde u}_2 - {\tilde u}_1 ) \ , \qquad 
{\tilde \alpha}_2 = 1 - {\tilde \alpha}_1 \ .
\label{tildealpha}
\end{eqnarray}
The plus sign in (\ref{tildeu21}) corresponds to
${\tilde u}_2$. For simplicity of notation, we omitted the
superscripts in the coefficients $f_i^{(1)}\!\equiv\!f_i(Q_1^2)$.
Note that expressions (\ref{f11f12})--(\ref{f13}) should be
inserted into (\ref{tildeu21})--(\ref{tildealpha}) in order
to obtain these parameters explicitly in terms of the
original coefficients $r^{(1)}_i\!\equiv\!r_i(Q_1^2)$
($i\!=1,\!2,\!3$). Inserting now these parameters
into expression (\ref{dBGAdef}) ($M\!=\!2$) gives the
sought for RScl--invariant approximation to $S^{(1)}_3$.
Several subcases should be distinguished (for $M\!=\!2$):
\begin{enumerate}
\item
When $(f_2\!-\!f_1^2)\!>\!0$, then:
${\tilde u}_i,\!{\tilde \alpha}_i$ are real ($i\!=\!1,\!2$),
${\tilde u}_1\!\not=\!{\tilde u}_2$ and 
$0\!<\!{\tilde \alpha}_i\!<\!1$.
\item
When $(f_2\!-\!f_1^2)\!<\!0$ and
$|f_3\!+\!f_1(2 f_1^2\!-\!3 f_2)|\!>\!2 \sqrt{ (f_1^2\!-\!f_2)^3 }$,
then: 
${\tilde u}_i,\!{\tilde \alpha}_i$ are real ($i\!=\!1,\!2$) and
${\tilde u}_1\!\not=\!{\tilde u}_2$.
\item
When $(f_2\!-\!f_1^2)\!<\!0$ and
$|f_3\!+\!f_1(2 f_1^2\!-\!3 f_2)|\!<\!2 \sqrt{ (f_1^2\!-\!f_2)^3 }$,
then: 
${\tilde u}_i$ are complex, ${\tilde \alpha}_i$
generally complex ($i\!=\!1,\!2$),\footnote{
It can be checked that also in this case the modified dBGA
(\ref{dBGAdef}) is real, 
because (\ref{tildeu21})-(\ref{tildealpha})
imply: ${\tilde u}_2\!=\!({\tilde u}_1)^{\ast}$ and
${\tilde \alpha}_2\!=\!({\tilde \alpha}_1)^{\ast}$.}
and ${\tilde u}_1\!\not=\!{\tilde u}_2$.
\item
When $(f_2\!-\!f_1^2)\!<\!0$ and
$|f_3\!+\!f_1(2 f_1^2\!-\!3 f_2)|\!=\!2 \sqrt{ (f_1^2\!-\!f_2)^3 }$
[or when $(f_2\!-\!f_1^2)\!=\!0$ and $f_3\!\not=\!f_1^3$], then:
the system of equations for ${\tilde u}_i$ and ${\tilde \alpha}_i$ is
not solvable, i.e., form (\ref{PAF1decomp}) is not valid,
the dPA (\ref{PAF11}) has a multiple (double) pole.
\item
When $(f_2\!-\!f_1^2)\!=\!0$ and $f_3\!=\!f_1^3$, then:
${\tilde u}_i,\!{\tilde \alpha}_i$ are real ($i\!=\!1,\!2$) and
${\tilde u}_1\!=\!{\tilde u}_2\!=\!f_1$.
\end{enumerate}

We should note that in QCD, presently available results
of perturbative calculations include for various observables
$S$ the coefficients $r_1^{(1)}$ and $r_2^{(1)}$ of (\ref{S}),
but not yet $r_3^{(1)}$. Therefore, at this stage, the
algorithm described here still cannot be applied for
$M\!=\!2$ in the case of QCD observables
($r^{(1)}_{2M\!-\!1}\!\equiv\!r^{(1)}_3$ are not available yet) .
This contrasts with QED where perturbative coefficients 
$r_3^{(1)}$ have been obtained for several QED observables.

One may raise the question of how to construct, e.g.~in QCD, 
an explicit approximate expression for
the function $k(z_1,u_1)$ appearing in the
modified dBGA's (\ref{dBGAdef}),
once we go beyond the large--${\beta}_0$ approximation.
In the case of QCD (or QED), 
$k(z_1,u_1)$ is defined in (\ref{comm1Theor2}).
For example, the QCD coupling parameter 
$z_1\!\equiv\!{\alpha}_s(Q_1^2)/\pi$
at two--loop level and written as an expansion in inverse
powers of $\ln Q_1^2$, is given in \cite{PDB}\footnote{
Note that convention (\ref{alphaRGE}) for 
QCD ${\beta}_i$ coefficients 
here is different from that in Particle Data Book
\cite{PDB}: ${\beta}_0^{PDB}\!=\!4{\beta}_0\!=\!11\!-\!2 n_f/3$;
${\beta}_1^{PDB}\!=\!8{\beta}_1\!=\!51\!-\!19 n_f/3$;
${\beta}_2^{PDB}\!=\!128{\beta}_2\!=\!2857\!-\!5033 
n_f/9\!+\!325 n_f^2/27$. Here, 
$n_f$ is number of effective quark flavors with mass less than
the considered scale $Q_1$.
The form for ${\beta}_2$ is in
${\overline {\mbox{MS}}}$ scheme; ${\beta}_0$ and ${\beta}_1$
are RSch--independent.}
\begin{equation}
z_1 \equiv \frac{ {\alpha}_s(Q_1^2) }{\pi} =
\frac{1}{ {\beta}_0 \ln ( Q_1^2/ {\Lambda}^2 ) }
\left[ 1 - \frac{ {\beta}_1 }{ {\beta}_0^2 }
\frac{ \ln \left( \ln \frac{Q_1^2}{ {\Lambda}^2 } \right) }
{ \ln \frac{Q_1^2}{ {\Lambda}^2 } } \right] \ ,
\label{alpha2l}
\end{equation}
where the neglected terms are of order 
${\beta}_1^2 \ln^2 [\ln(Q_1^2/{\Lambda}^2)]/\ln^3(Q_1^2/{\Lambda}^2)$,
and ${\Lambda}$ is the QCD scale which
depends (like ${\beta}_i$'s) on the number of effective quark flavors
$n_f$, e.g., ${\Lambda}(n_f\!=\!5)\!\approx\!200$ MeV,
${\Lambda}(n_f\!=\!4)\!\approx\!280$ MeV.
It should be stressed that 
${\Lambda}$ is renormalization--scheme--dependent
(RSch--dependent), hence change of RSch is
equivalent to change of RScl when effects of ${\beta}_j$'s 
($j\!\geq\!2$) are neglected.
In order to find $k(z_1,u_1)$ [note: $u_1\!=\!\ln(p^2/Q_1^2)$],
we can first write (\ref{alpha2l}) for the scale $p^2$
\begin{equation}
z_1 k(z_1,u_1) \equiv \frac{ {\alpha}_s(p^2) }{\pi} =
\frac{1}{ {\beta}_0 \left[ \ln ( Q_1^2/ {\Lambda}^2 ) + u_1 \right] }
\left\{  1 - \frac{ {\beta}_1 }{ {\beta}_0^2 }
\frac{ \ln \left[ \ln \left( \frac{Q_1^2}{ {\Lambda}^2 } \right)
+ u_1 \right] }
{ \left[ \ln \left( \frac{Q_1^2}{ {\Lambda}^2 } \right) + u_1 \right] }
\right\} \ .
\label{alphak22l}
\end{equation}
To obtain an explicit form for $k(z_1,u_1)$, we
express $\ln ( Q_1^2/{\Lambda}^2 )$ in (\ref{alphak22l})
by $z_1$ via (\ref{alpha2l}). Inverting
(\ref{alpha2l}) can be done numerically, for example
by iteration
\begin{equation}
\left[ \ln \frac{Q_1^2}{ {\Lambda}^2 } \right]^{(n+1)}
= \frac{1}{ {\beta}_0 z_1 } \left\{ 1 -
\frac{ {\beta}_1 }{ {\beta}_0^2 }
\frac{ \ln \left[ \ln \frac{Q_1^2}{ {\Lambda}^2 } \right]^{(n)} }
{ \left[ \ln \frac{Q_1^2}{ {\Lambda}^2 } \right]^{(n)} } 
\right\} \ , \qquad
\left[ \ln \frac{Q_1^2}{ {\Lambda}^2 } \right]^{(0)} =
\frac{1}{ {\beta}_0 z_1 } \ .
\label{iter12}
\end{equation}
If we choose to stop already after the first iteration step,
we get
\begin{equation}
\left[ \ln \frac{Q_1^2}{ {\Lambda}^2 } \right]^{(1)} =
\frac{1}{ {\beta}_0 z_1 } + \frac{ {\beta}_1 }{ {\beta}_0^2 }
\ln ( {\beta}_0 z_1 ) \ ,
\label{iter3}
\end{equation}
\begin{eqnarray}
\lefteqn{
k(z_1,u_1) \left[ \equiv \frac{ {\alpha}_s(p^2) }{ {\alpha}_s(Q_1^2) }
\right] = \frac{1}{ \left[ 1 + {\beta}_0 z_1 u_1
+  ({\beta}_1/{\beta}_0) z_1 \ln ( {\beta}_0 z_1 ) \right] }
{\Bigg \{} 1 + } 
\nonumber\\
&&+ \frac{  ({\beta}_1/{\beta}_0) z_1
\left[ \ln \left( {\beta}_0 z_1 \right) - 
\ln \left( 1 + {\beta}_0 z_1 u_1 
+ ({\beta}_1/{\beta}_0) z_1 \ln ( {\beta}_0 z_1 ) \right) \right] }
{ \left( 1 + {\beta}_0 z_1 u_1 
+ ({\beta}_1/{\beta}_0) z_1 \ln ( {\beta}_0 z_1 ) \right) }
{\Bigg \}} \ .
\label{kz1u1iter}
\end{eqnarray}
This would then be a $k(z_1,u_1)$ function beyond
large-${\beta}_0$ approximation that could be used in the
RScl-invariant expressions for modified dBGA's (\ref{dBGAdef}),
with parameters ${\tilde u}_i$ and ${\tilde \alpha}_i$
in the case of $M\!=\!2$ determined from the original
truncated series $S^{(1)}_3$ by Eqs.~(\ref{f11f12})--(\ref{tildealpha}).
In the latter equations, ${\beta}_2\!$ would have to be set equal
to zero, because (\ref{alpha2l})--(\ref{kz1u1iter}) represent
solutions of RGE (\ref{alphaRGE}) with
only ${\beta}_0$ and ${\beta}_1$ retained. The
RScl--invariance of the obtained modified dBGA in $M\!=\!2$ case
would still not be absolutely precise because
(\ref{alpha2l}) and (\ref{kz1u1iter}) represent only
an {\em approximate} solution to truncated 
(${\beta}_2\!=\!{\beta}_3\!=\!\ldots\!=\!0$) RGE (\ref{alphaRGE}).
To make RScl--invariance precise, we would have to
integrate the latter truncated RGE numerically from
$t\!\equiv\!\ln(p^2/Q_1^2)\!=\!0$ to $t\!=\!{\tilde u}_i$
($i\!=\!1,\!2$) to obtain $k(z_1,{\tilde u}_i)$, 
provided we have at RScl $Q_1^2$ a
reasonably reliable value of 
${\alpha}_s(Q_1^2)\!\equiv\!{\pi} z_1$.

In fact, numerical integration of the RGE (\ref{alphaRGE})
would always work and give us such $k(z_1,{\tilde u}_i)$'s
in (\ref{dBGAdef})
that RScl--invariance would be precise. In this RGE, we should
not neglect at least those ${\beta}_j$ coefficients which
appear as nonzero numbers in the expressions for the coefficients
$f^{(1)}_j$ ($j\!=\!1,\!\ldots,\!2M\!-\!1$) of the rearranged
series (\ref{Sinkj}), e.g.~for $M\!=\!2$ case we should
not neglect ${\beta}_1$ and ${\beta}_2$ 
[cf.~Eqs.~(\ref{f11f12})--(\ref{f13})].
On the other hand, we can take into account in the RGE (\ref{alphaRGE})
more than these minimally required coefficients ${\beta}_j$,
thus making the result (\ref{dBGAdef}) RScl--invariant at such
an improved level.
We note, however, that only a limited number of perturbative
coefficients ${\beta}_j$ (namely: 
${\beta}_0,\!\ldots,\!{\beta}_3$) are known in QCD
(cf.~\cite{QCDbeta}, in ${\overline {\mbox{MS}}}$ scheme) and 
QED (cf.~\cite{QEDbeta}, in ${\overline {\mbox{MS}}}$, MOM and
in on-shell schemes). Therefore, the RGE (\ref{alphaRGE})
will always be truncated at some level. 
 
An alternative way to obtain $k(z_1, {\tilde u}_i)$, for
known $z_1$ and ${\tilde u}_i$, would be to solve (a truncated)
RGE (\ref{alphaRGE}) by integrating it analytically from 
$t\!=\!t_1\!\equiv\!\ln(Q_1^2/Q_1^2)\!=\!0$ to $t\!=\!{\tilde u}_i$,
beforehand Taylor--expanding the
integrands on the right around the point $t\!=\!0$
and repeatedly using the same RGE (\ref{alphaRGE})
in evaluating Taylor coefficients.
If we ignore threshold effects for simplicity,
this procedure leads to the following expression:
\begin{equation}
k(z_1,{\tilde u}_i) = 1 + z_1 ( - {\beta}_0 {\tilde u}_i )
+ z_1^2 ( {\beta}_0^2 {\tilde u}_i^2 - {\beta}_1 {\tilde u}_i )
+ z_1^3 ( - {\beta}_0^3 {\tilde u}_i^3 +
\frac{5}{2} {\beta}_0 {\beta}_1 {\tilde u}_i^2 - 
{\beta}_2 {\tilde u}_i ) + \ldots \ .
\label{kfunexpan}
\end{equation}
If we take into account only a limited number of
${\beta}_i$ coefficients, than (\ref{kfunexpan})
may represent a convergent, or a well behaved
asymptotic, series for small enough
parameters ${\tilde u}_i$. We conjecture that
the latter (generally complex) parameters
have reasonably small absolute values if our choice of
$Q_1^2$ is judicious, i.e., if $Q_1^2$ is close to a typical scale of
the process considered.

\section{Conclusions, open questions}
We presented a method of constructing approximants to available
truncated perturbative series (TPS's) of QED or QCD
observables. The approximants are modified diagonal Baker--Gammel
approximants (dBGA's), with the kernel being a ratio of
the gauge coupling parameters at different scales. 
We showed that these approximants have
two favorable properties: they reproduce the
TPS to the available order when expanded in powers of the
gauge coupling parameter, and they are invariant under
the change of the renormalization scale (RScl). The
gauge coupling parameter can be taken to evolve with
the scale at {\em any\/} chosen loop-order.
 
There are several questions that the presented work raises.
For example, it remains unclear whether the dPA's (\ref{PAF11})\
could have multiple poles in some exceptional physical cases,
and if so, what would have to be changed in the presented formalism
in such a case. Furthermore, the formalism offers
RScl--invariant approximants to truncated perturbation series
$S^{(1)}_n$ of (\ref{Sn}) only for odd $n\!=\!2M\!-\!1$. What would
be the best approximant in cases of even $n\!=\!2M$ (especially
in case $n\!=\!2$)\footnote{
The presented algorithm probably cannot be trivially extended
to the cases of $n\!=\!2M$, because it relies heavily on the
decomposition (\ref{PAF1decomp}) which is valid only for
{\em diagonal\/} PA's.}?
This question remains open even
in the large--${\beta}_0$ approximation when
BGA's reduce to the usual PA's.
The question of when the series of modified dBGA's
{$ z_1 G_{ f^{(1)} }^{ [M-1/M] }(z_1) $} ($M\!=\!1,\!2,\!\ldots$),
defined by (\ref{dBGAdef}),
is  convergent or asymptotic also remains open,
especially since our proofs of Theorems 1 and 2 implicitly
assumed convergence of series involved in the proofs. The latter
question appears also when dealing with the usual dPA's -- however,
in such a case a lot of progress in answering this question has
already been achieved -- from formal mathematical
\cite{BakerMorris} and from physical/empirical
point of view \cite{SEK}. We also note that the dBGA result 
(\ref{dBGAdef}) becomes at the level $M\!\geq\!2$
in principle RSch--dependent, since parameters
in the dBGA involve also ${\beta}_2$ 
(and possibly higher ${\beta}_j$'s) coefficients 
[cf.~(\ref{f11f12})--(\ref{tildealpha})] which
are explicitly scheme--dependent. Therefore, it would
be instructive to test numerically RSch--dependence of
dBGA (\ref{dBGAdef}) for $M\!\geq\!2$ cases. We expect that 
this RSch--dependence (-- dependence on ${\beta}_2$) would
be very weak, at least for such choices of RScl $Q_1^2$ for which
$z_1\!\equiv\!{\alpha}_s(Q_1^2)/{\pi}$ is not
large ($z_1\!\stackrel{<}{\sim}\!0.09$),
since work of Ref.~\cite{SEK} (third entry
-- PRD) indicates that this is the case for some PA's 
when observable $S$ is the effective charge in the Bjorken sum rule.
In any case, one of the next
natural steps would be to study efficiency of the presented formalism
in cases of actual available truncated perturbative series for QED
observables, to compare it numerically with other methods,
and to investigate which
classes of Feynman diagrams the presented method actually
sums up. 

Furthermore, the method, with some modifications, 
may prove useful also in areas
of physics others than high energy physics.

\vspace{1cm}

\noindent {\bf Acknowledgements:\/}

The author wishes to thank Professors D.~Schildknecht and 
R.~K\"ogerler for offering him financial support of Bielefeld
University during the course of this work.

\vspace{1.cm}

\noindent {\footnotesize
\bf Abbreviations used frequently in the article\/:}\\
{\footnotesize (d)BGA -- (diagonal) Baker--Gammel approximant; 
(d)PA -- (diagonal) Pad\'e approximant; 
RSch -- renormalization scheme; 
RScl -- renormalization scale;
TPS -- truncated perturbation series.}

\end{document}